\documentclass[showpacs,amsmath,amssymb,reprint,aip,apl]{revtex4-1}

\usepackage{graphicx}
\usepackage{dcolumn}
\usepackage{bm}
\usepackage{color}

\begin{document}

\preprint{}

\title{Fine tuning epitaxial strain in ferroelectrics: Pb$_x$Sr$_{1-x}$TiO$_3$ on DyScO$_3$}
\date{\today}

\author{G. Rispens}
\altaffiliation{Present adress: DPMC, University of Geneva, Quai Ernest Ansermet 24, CH-1211 Gen\`{e}ve 4, Switzerland}
\author{J.A. Heuver}
\author{B. Noheda}
\affiliation{Zernike Institute for Advanced Materials, University of Groningen, Nijenborgh 4, 9747 AG Groningen, The Netherlands}


\begin{abstract}Epitaxial strain can be efficiently used to modify the properties of ferroelectric thin films. From the experimental viewpoint, the challenge is to fine-tune the magnitude of the strain. We illustrate here how, by using a suitable combination of composition and substrate, the magnitude of the epitaxial strain can be controlled in a continuous manner. The phase diagram of Pb$_x$Sr$_{1-x}$TiO$_3$ films grown epitaxially on (110)-DyScO$_3$ is calculated using a Devonshire-Landau approach. A boundary between an in-plane and an out-of-plane oriented ferroelectric phases is predicted to take place at $x\approx 0.8$. A series of Pb$_x$Sr$_{1-x}$TiO$_3$ epitaxial films grown by Molecular Beam Epitaxy show good agreement with the proposed phase diagram. 
\end{abstract}

\maketitle

\noindent Modifying the properties of crystalline thin films using the epitaxial strain induced by the substrate as an adjustable parameter is known as \emph{strain engineering} or \emph{strain tuning}\cite{Schlom2007}. This has captured the interest of several condensed matter communities because ordering temperatures can be increased, physical responses can be enhanced and even new functionalities can be added to thin films utilizing epitaxial strain\cite{Broe1991,Loc1998,Jai2000,Ram2007}. 

Ferroelectric materials are especially suitable for such strain studies due to their strong coupling between strain and electrical polarization.
However, despite some very impressive experimental results\cite{Bea2009,Biegalski2006,Haeni2004,Warusawithana2009,Zeches2009,Lee2010}, the realization of strain engineering in ferroelectrics lags behind the predictions and it is difficult to fully employ the wealth of interesting possibilities suggested by the theorists\cite{Bousquet2010,Dieguez2005,Ederer2005,Fennie2006,Koukhar2001a,Pertsev2000,Pertsev1998}. The limited number of suitable substrate materials is an important factor in this. Two substrates that have a good lattice match with many functional perovskites are SrTiO$_3$ and DyScO$_3$, but there is more than a percent difference between their (pseudo)cubic lattice parameters\cite{Biegalski2005}(see Figure 1a). This difference is too large if one aims to establish an experimental temperature-strain (T-u) phase diagram. Moreover, a too large mismatch will induce various relaxation mechanisms that will prevent elastic strain accommodation.\cite{Speck1994a,Vlooswijk2007}. Most importantly, only if the strain can be tuned continuously, we will be able to access the novel phases\cite{Pertsev1998} or novel properties\cite{Iniguez2010,Damjanovic2010} that are theoretically predicted, and which exist only for a narrow region of strain values. 

\begin{figure}
	\centering
		\includegraphics[width=0.49\textwidth]{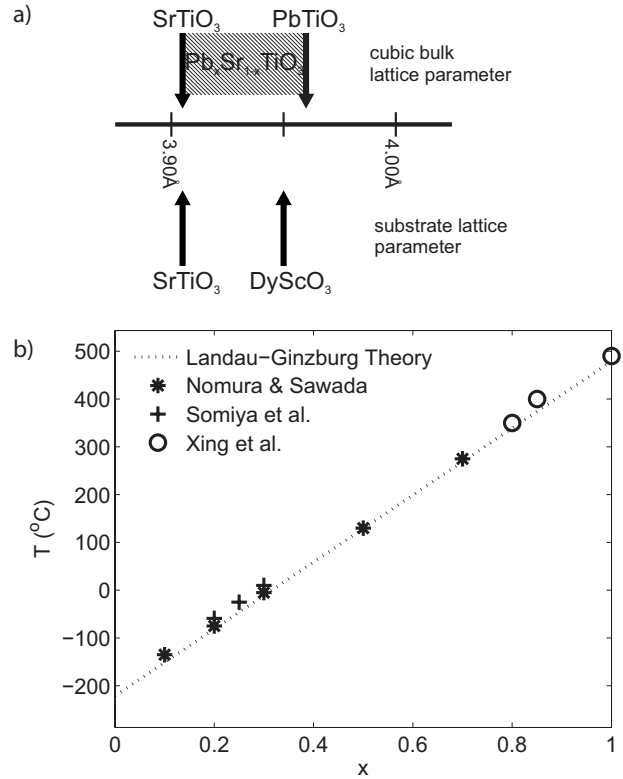}
	\caption{a) Comparison of the lattice parameter of SrTiO$_3$ and DyScO$_3$ substrates and Pb$_x$Sr$_{1-x}$TiO$_3$ thin films. b) Transition temperature of Pb$_x$Sr$_{1-x}$TiO$_3$ versus the Pb content $x$. The solid line is calculated from LD theory. The data points are taken from various literature sources\cite{NOMURA1955,Somiya2001,Xing2003}.}
	\label{fig:PST_intro}
\end{figure}

In this letter we combine the epitaxial strain imposed by the substrate with compositional variations of the film in order to change the magnitude of the strain in a continuous manner. For that we use Sr-substituted PbTiO$_3$ thin films grown on DyScO$_3$ substrates. Various reasons led us to choose these materials. The lattice parameters of Pb$_x$Sr$_{1-x}$TiO$_3$, as well as the Curie temperatures, T$_C$, vary linearly between the two end members of the solid solution \cite{NOMURA1955,Somiya2001,Xing2003}(see figure \ref{fig:PST_intro}b). Moreover, above room temperature, the bulk solid solution does not show other phases than the well-known paraelectric cubic and ferroelectric tetragonal phase of PbTiO$_3$\cite{Xing2003}. In this way we can use a phenomenological Landau-Devonshire (LD) approach to calculate the phase diagram of Pb$_x$Sr$_{1-x}$TiO$_3$ epitaxial thin films on (110)-DyScO$_3$. The substrate was chosen because the strain state of Pb$_x$Sr$_{1-x}$TiO$_3$ films epitaxially grown on (110)-DyScO$_3$ can be changed from (slightly) compressive to tensile by varying the Sr-content. This is in agreement with other reports showing that the polarization of epitaxial PbTiO$_3$ films on (110)-DyScO$_3$ is predominantly out-of-plane\cite{Catalan2006}, whereas the polarization of epitaxial SrTiO$_3$ films on DyScO$_3$ is in the plane of the film\cite{Biegalski2006,Biegalski2009,Haeni2004}. 

\begin{figure}[ht]
	\centering
		\includegraphics[width=0.49\textwidth]{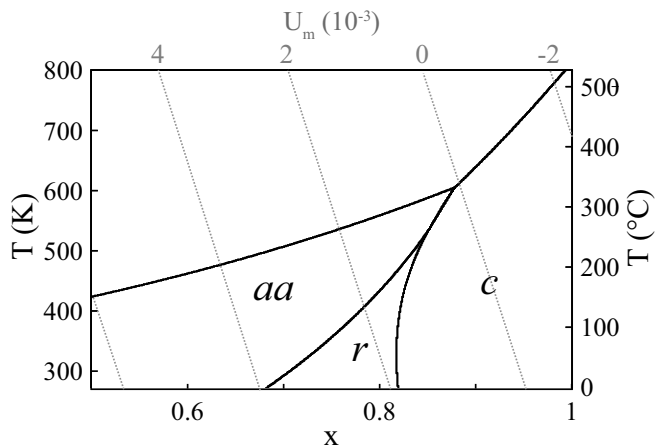}
	\label{fig:PDPST-DSO2}
	\caption{Phase diagram versus composition for Pb$_x$Sr$_{1-x}$TiO$_3$ strained on DyScO$_3$ as calculated using LD theory. The top axis defines the 'iso-strain' lines in the composition versus temperature diagram.}
\end{figure}

The phase diagram of Pb$_x$Sr$_{1-x}$TiO$_3$ on (110)-DyScO$_3$ (110) was calculated including the epitaxial strain in the LD free energy expansion, as described by Pertsev et al.\cite{Pertsev1998}. The composition-dependent Landau coefficients of Pb$_x$Sr$_{1-x}$TiO$_3$ have been constructed as a linear combination of the well-known Landau coefficients of the end members, PbTiO$_3$ and SrTiO$_3$, similar to refs.\cite{Ban2002,Dawber2007,Shir2009}. The Landau coefficients of the end members are those of refs. \onlinecite{HAUN1987} and \onlinecite{Shir2009}, respectively. \footnote{all values are given in SI units (T = K, a$_1$ = J m C$^{-2}$, a$_{ij}$ = J m$^5$ C$^{-4}$, a$_{ijk}$ = J m$^9$ C$^{-6}$, Q$_{ij}$ = m$^4$ C$^{-2}$, s$_{ij}$ = m$^3$ J$^{-1}$). For PbTiO$_3$: a$_1$ = 3.8$\times$10$^5$(T-752), a$_{11}$ = -7.3$\times$10$^7$, a$_{12}$ = 7.5$\times$10$^8$, a$_{111}$ = 2.6$\times$10$^8$, a$_{112}$ = 6.1$\times$10$^8$, a$_{123}$ = -3.7$\times$10$^9$, Q$_{11}$ = 0.089, Q$_{12}$ = -0.026, Q$_{44}$ = 0.0675, s$_{11}$ = 3.7$\times$10$^{-12}$, s$_{12}$ = -2.5$\times$10$^{-12}$, s$_{44}$ = 9.0$\times$10$^{-12}$. For SrTiO$_3$: a$_1$ = 7.06$\times$10$^5$(T-35.5) (this is an approximation only valid above 100K\cite{PhysicsFE}), a$_{11}$ = 1.04$\times$10$^8$, a$_{12}$ = 0.746$\times$10$^8$, a$_{111}$ = 0, a$_{112}$ = 0, a$_{123}$ =0, Q$_{11}$ = 0.0496, Q$_{12}$ = -0.0131, Q$_{44}$ = 0.019, s$_{11}$ = 3.52$\times$10$^{-12}$, s$_{12}$ = -0.85$\times$10$^{-12}$, s$_{44}$ = 7.87$\times$10$^{-12}$.}

For the temperature dependent coefficient $a_1= \alpha_1 (T-T_C)$, the composition dependence of T$_C$ and $ \alpha_1$ are treated separately. The misfit strain depends on both composition and temperature, because of differences in thermal expansion between film and substrate. As the thermal expansion for PbTiO$_3$ and SrTiO$_3$ are almost equal\cite{Landolt-Bornstein}, the thermal expansion of Pb$_x$Sr$_{1-x}$TiO$_3$ is assumed to be that of PbTiO$_3$. The thermal expansion of DyScO$_3$ does significantly differ from that of Pb$_x$Sr$_{1-x}$TiO$_3$ \cite{Biegalski2005}. The small anisotropy of 0.05\% in the lattice parameters of the (110) DyScO$_3$ plane (3.945$\AA$ vs 3.947$\AA$ at room temperature) was neglected, and the average of the $a$ and $b$ lattice parameters was used as the in-plane lattice parameter in the calculations. This is justified by the results in ref. \cite{Zembilgotov2005} showing that Landau simulations on single domain PbTiO$_3$ and Pb$_{0.35}$Sr$_{0.65}$TiO$_3$ films give no qualitative difference in the phase diagram after including a substrate anisotropy as small as that of DyScO$_3$\cite{Zembilgotov2005}. Since the oxygen rotations present in SrTiO$_3$ below 105K are not included in the calculations, our results are not expected to be valid for x $<$ 0.5\cite{Shir2009}.  Finally, the LD approach used here considers uniform polarization throughout the film, thus possible domain formation and polarization gradients are not taken into account in this approximation. The resulting free energy expansion was minimized with respect to the Cartesian components of the polarization (along the axes of the perovskite unit cell) to obtain a phase diagram as a function of Pb content ($x$).

\begin{figure*}
	\centering
		\includegraphics[width=0.98\textwidth]{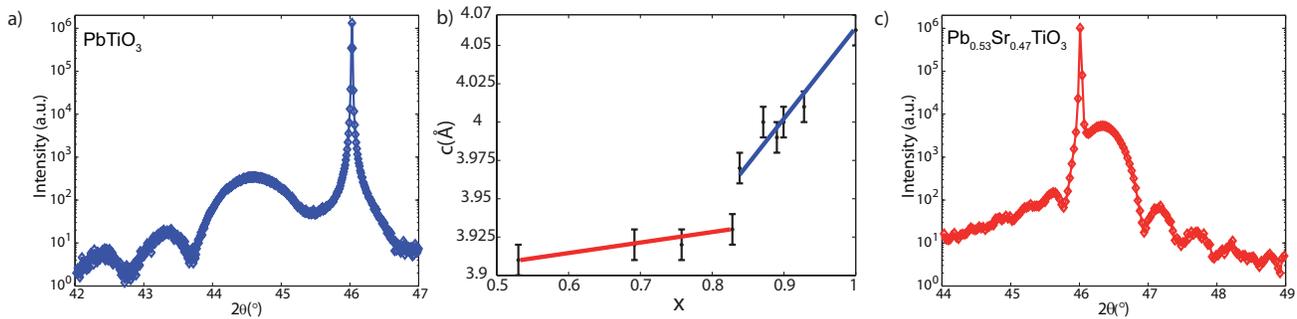}
	\caption{Out-of-plane XRD results for Pb$_x$Sr$_{1-x}$TiO$_3$ films grown on DyScO$_3$. The middle graph shows the evolution of the $c$ lattice parameter with $x$. There is a clear transition with decreasing $x$ from a unit cell with a long out-of-plane lattice parameter to a unit cell with  long in-plane lattice parameters.}
	\label{fig:PSTDSOCvsX}
\end{figure*}

The calculated phase diagram of Pb$_x$Sr$_{1-x}$TiO$_3$ on DyScO$_3$ is shown in figure \ref{fig:PDPST-DSO2}. The misfit strain goes from a very small compressive strain for pure PbTiO$_3$ to a tensile strain that increases with increasing Sr content. The different strain values stabilize polarizations along different directions. For large $x$, Pb$_x$Sr$_{1-x}$TiO$_3$ is predicted to be a $c$-phase ferroelectric, with the electrical polarization, P, perpendicular to the film plane, similar to that of PbTiO$_3$. At lower $x$ an $aa$-phase, with P $\|$ $\left\langle 110 \right\rangle$, should be stabilized. In between these two, an intermediate $r$-phase is expected, in which the polarization points somewhere in between $\left\langle 001 \right\rangle$ and $\left\langle 110 \right\rangle$. The addition of Sr gives rise to the decrease in T$_c$. 

To test our predictions, a series of Pb$_x$Sr$_{1-x}$TiO$_3$ thin films with a thickness of 50 monolayers (~20nm) were grown on (110)-DyScO$_3$ using Molecular Beam Epitaxy (MBE). The films were grown at a substrate temperature of 650$^o$C, with an adsorption controlled growth mechanism as described for PbTiO$_3$\cite{Theis1997,Theis1998,Rispens2007}. Sr substitution is obtained by providing a constant flux of atomic Sr for a certain amount of time, t$_{Sr}$, at each monolayer. Figure \ref{fig:PSTDSOCvsX} shows the out-of-plane lattice parameter $c$, obtained from XRD 2$\theta$-$\omega$ scans, versus the Pb content, $x$. At large $x$, a lattice parameter larger than the pseudo-cubic lattice parameters of DyScO$_3$ (c= 3.945 \AA) is observed in the films. For fully strained films and neglecting the small difference between the two in-plane lattice parameters of the substrate, this leads to a tetragonal structure similar to that of bulk PbTiO$_3$. The polarization is expected to be along the symmetry axis, so this corresponds to a $c$-ferroelectric phase\cite{Pertsev1998}, with P$\|$[001]. At x$\approx$ 0.83, there is a discontinuous decrease in the out-of-plane lattice parameter to a value smaller than that of DyScO$_3$, leading to a structure with larger in-plane lattice parameters. Here the polarization is expected to lie in the plane of the film.

The experimental observations are in good agreement with the proposed phase diagram. In particular, the boundary between the in-plane and out-of-plane polarization is well reproduced experimentally. Temperature-dependent measurements indicate that the Curie temperatures of the strained films are also in agreement with the calculated ones\cite{Rispens2011}. However, the assumption of uniform polarization throughout the film is most likely not valid, since domains are expected to form \cite{Catalan2006,Biegalski2009}, most likely modifying the phase diagram. Reciprocal space maps in our films reveal, indeed, the presence of domains\cite{Rispens2011}. Next we plan to look into the mechanisms of domain formation as well as to grow similar films with bottom and top electrodes in order to investigate the ferroelectric properties across the phase diagram.

In summary we have shown that the misfit strain in ferroelectric thin films can be fine-tuned by using a suitable combination of composition and substrate. This strategy was applied to epitaxial Pb$_x$Sr$_{1-x}$TiO$_3$ films on (110)-DyScO$_3$. The calculated phase diagram as a function of composition (which implies changes in epitaxial strain) predicts a phase landscape similar to that in the phase diagram of PbTiO$_3$ as a function of strain\cite{Pertsev1998} (in-plane, out-of-plane and intermediate polar phases). In the present case, continuous tuning across the phase diagram can be achieved. The structural evolution of a series of Pb$_x$Sr$_{1-x}$TiO$_3$ epitaxial films is consistent with a change in the polarization direction, in good agreement with the proposed phase diagram. 

We are grateful to Sergey Artyukhin for useful discussions. This work is part of the research programme of the Foundation for Fundamental Research on Matter (FOM), which is part of the Netherlands Organisation for Scientific Research (NWO).

\end {document}